\documentclass[]{mnras}
\usepackage{graphicx}	
\usepackage{amsmath}	
\usepackage{amssymb}	
\usepackage{multicol}        
\usepackage{bm}		
\usepackage{pdflscape}	
\usepackage{subfig}
\usepackage{newtxtext,newtxmath}
\usepackage{gensymb}
\usepackage[switch]{lineno}
 
\usepackage[T1]{fontenc}
\usepackage{ae,aecompl}

\usepackage{newtxtext,newtxmath}

\title[Feedback and Feeding in Centaurus A]{Multi-scale feedback and feeding in the closest radio galaxy Centaurus A}

\author[B. McKinley et al.]
{B.~McKinley,$^{1}$ 
S.~J.~Tingay,$^{1}$
M.~Gaspari,$^{2,3}$
R.~P.~Kraft,$^{4}$
C.~Matherne,$^{5}$
A.~R.~Offringa,$^{6}$
M.~McDonald,$^{7}$
M.~S.~Calzadilla,$^{7}$
\newauthor
S.~Veilleux,$^{8}$
S.~S.~Shabala,$^{9,10}$
S.~D.~J.~Gwyn,$^{11}$
J.~Bland-Hawthorn,$^{12}$
D.~Crnojevi\'c,$^{13}$
B.~M.~Gaensler$^{14}$
M.~Johnston-Hollitt,$^{15}$
}

\begin{document}
\label{firstpage}
\maketitle

\begin{figure*}
\includegraphics[clip,trim=30 235 35 5,width=0.87\textwidth]{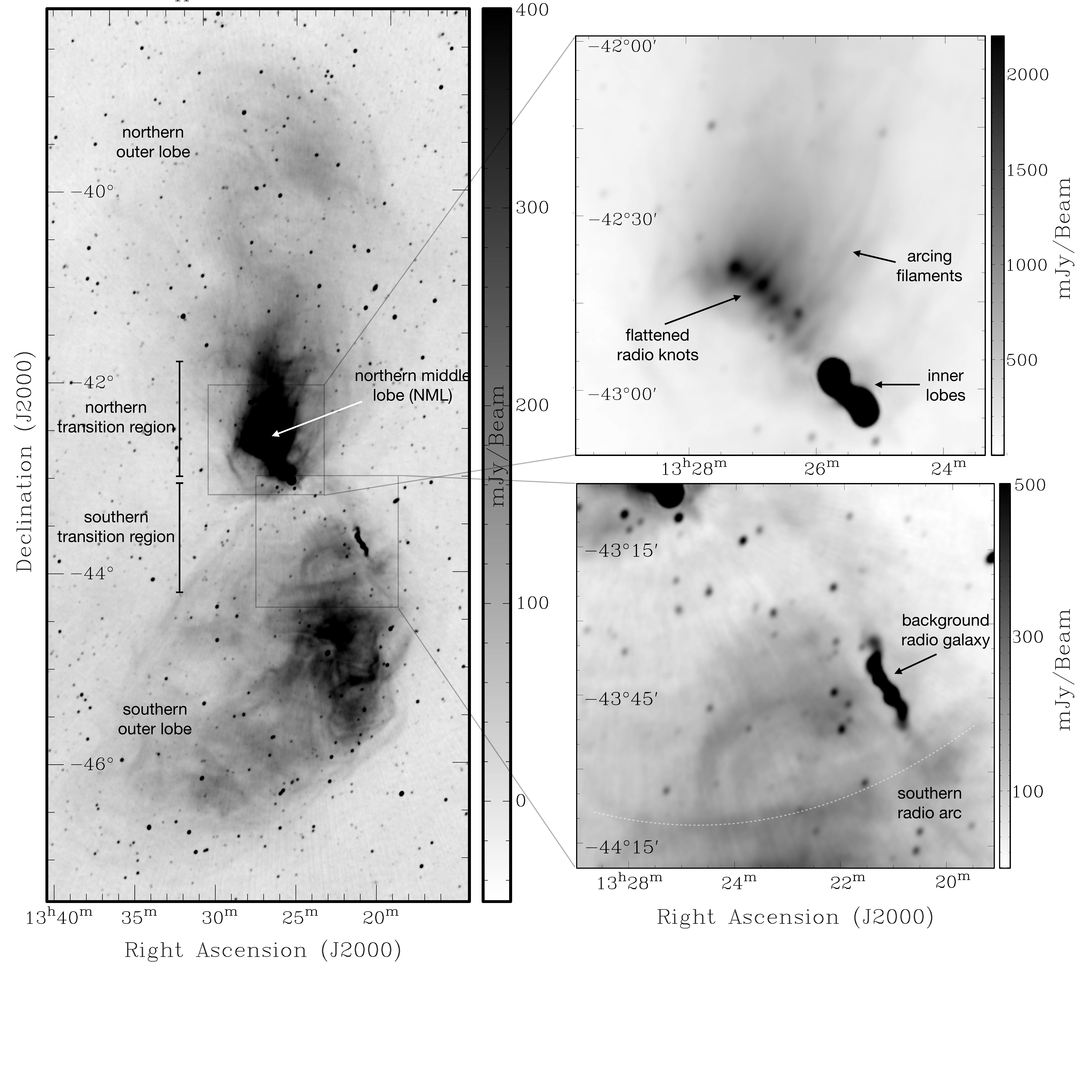}
\caption{Our nearest neighbouring radio galaxy, Centaurus~A, as seen at 185~MHz with the Murchison Widefield Array. Left: The whole radio source shown on a linear intensity scale between $-40$ and $400$~mJy/beam (the average restoring beam is a Gaussian of width 1.5$\times$1.2~arcmin with a major axis position angle of 155\degr). Right-top: The northern transition region on a linear intensity scale of between $-40$ and $2200$~mJy/beam, showing the flattened radio knots and arcing filaments. Right-bottom: The southern transition region on a linear intensity scale of between $0$ and $500$~mJy/beam, showing the southern radio arc. Each panel uses the same image data and no post-processing image manipulation has been performed.}
\label{fig:1}
\end{figure*}

\textbf{Supermassive black holes and supernovae explosions at the centres of active galaxies power cycles of outflowing and inflowing gas that affect galactic evolution and the overall structure of the Universe\cite{king2015,veilleux2020}. While simulations and observations show that this must be the case, the range of physical scales (over ten orders of magnitude) and paucity of available tracers, make both the simulation and observation of these effects difficult\cite{nelson2019,heckman2017}. By serendipity, there lies an active galaxy, Centaurus~A (NGC 5128)\cite{israel1998,feain2010}, at such a close proximity as to allow its observation over this entire range of scales and across the entire electromagnetic spectrum. In the radio band, however, details on scales of 10-100~kpc from the supermassive black hole have so far been obscured by instrumental limitations\cite{feain2011,neff2015a}. Here we report low-frequency radio observations that overcome these limitations and show evidence for a broad, bipolar outflow with velocity 1100~km s$^{-1}$ and mass outflow rate of 2.9 $M_{\odot}$ yr$^{-1}$ on these scales. We combine our data with the plethora of multi-scale, multi-wavelength historical observations of Centaurus~A to probe a unified view of feeding and feedback, which we show to be consistent with the Chaotic Cold Accretion self-regulation scenario\cite{gaspari2020,gaspari2013}.}

Centaurus~A has been studied in great detail across the entire electromagnetic spectrum and has the largest angular extent of any radio galaxy, due to its close proximity\cite{israel1998}. The centre of the galaxy is just 3.8$\pm$0.1 Mpc distant\cite{harris2010}, such that 1~arcmin on the sky equates to approximately 1~kpc. The radio source is characterised by a pair of bright inner lobes\cite{clarke1992}, which extend approximately 5~kpc from the nucleus and are being inflated by sub-pc-scale radio jets (see Fig.~\ref{fig:1}, for the locations of these key features). The 8$\degree$ outer radio lobes span at least 480~kpc, their true size likely being larger since the northern jet points toward us and the southern counterpart away from us\cite{tingay1998,EHT}, so the outer lobes probably also lie at an angle to our line-of-sight.

We used the Murchison Widefield Array (MWA\cite{tingay2013}) to observe the full extent of Centaurus~A at a central frequency of 185~MHz and with an angular resolution of 1.5 arcmin (see Methods), taking advantage of the MWA's extreme field-of-view, superb radio-quiet location and excellent sensitivity to large angular scales. The angular size and complexity of Centaurus~A have long posed challenges for the imaging capabilities of radio telescopes. The MWA is able to meet these challenges, providing unique insights into the nature of this complex object. We detect previously unseen features throughout the diffuse outer lobes of the radio galaxy, which consist of a complex network of filaments (see Fig.~\ref{fig:1}). Of particular interest are the areas $\sim1\degree$ (tens of kpc) north and south of the nucleus, which are transition regions through which energy must be transferred if feedback processes are able to significantly quench cooling flows, as proposed by theoretical models\cite{gaspari2011a,gaspari2017}.

With these transition regions revealed in our radio data, it is possible to study both the feeding and feedback processes occurring in Centaurus~A, across the full range of physical scales and using a multi-wavelength approach. Considering the available suite of data on Centaurus A, we find that the unified model of Active Galactic Nuclei (AGN) feeding and feedback\cite{gaspari2020,gaspari2017}, based on Chaotic Cold Accretion\cite{gaspari2013} (CCA) as the inflow mechanism, best explains the observations (see Methods for a detailed analysis and justification for the use of the unified CCA model). In this scenario, large-scale hot plasma halos cool and condense in a top-down multiphase cascade, forming clouds of cooler gas that rain down toward the nuclear region. A small amount of this matter is accreted, while most is ejected via fast outflows/jets that entrain further mass as they expand outwards with a larger opening angle. The model describes the link between the feeding of the AGN and the deposition of energy back into the surrounding medium via large-scale mechanical feedback and is agnostic to the source of the mechanical power. Wide slow outflows at large radii are therefore characteristic of the model, but narrower radio jets, such as those seen directly in Centaurus~A and other FRI type sources, can still be present and contribute to a wide large-scale feedback process\cite{krause2012}.

\begin{figure}
\centering 
\includegraphics[clip,trim=100 130 20 200,width=0.51\textwidth]{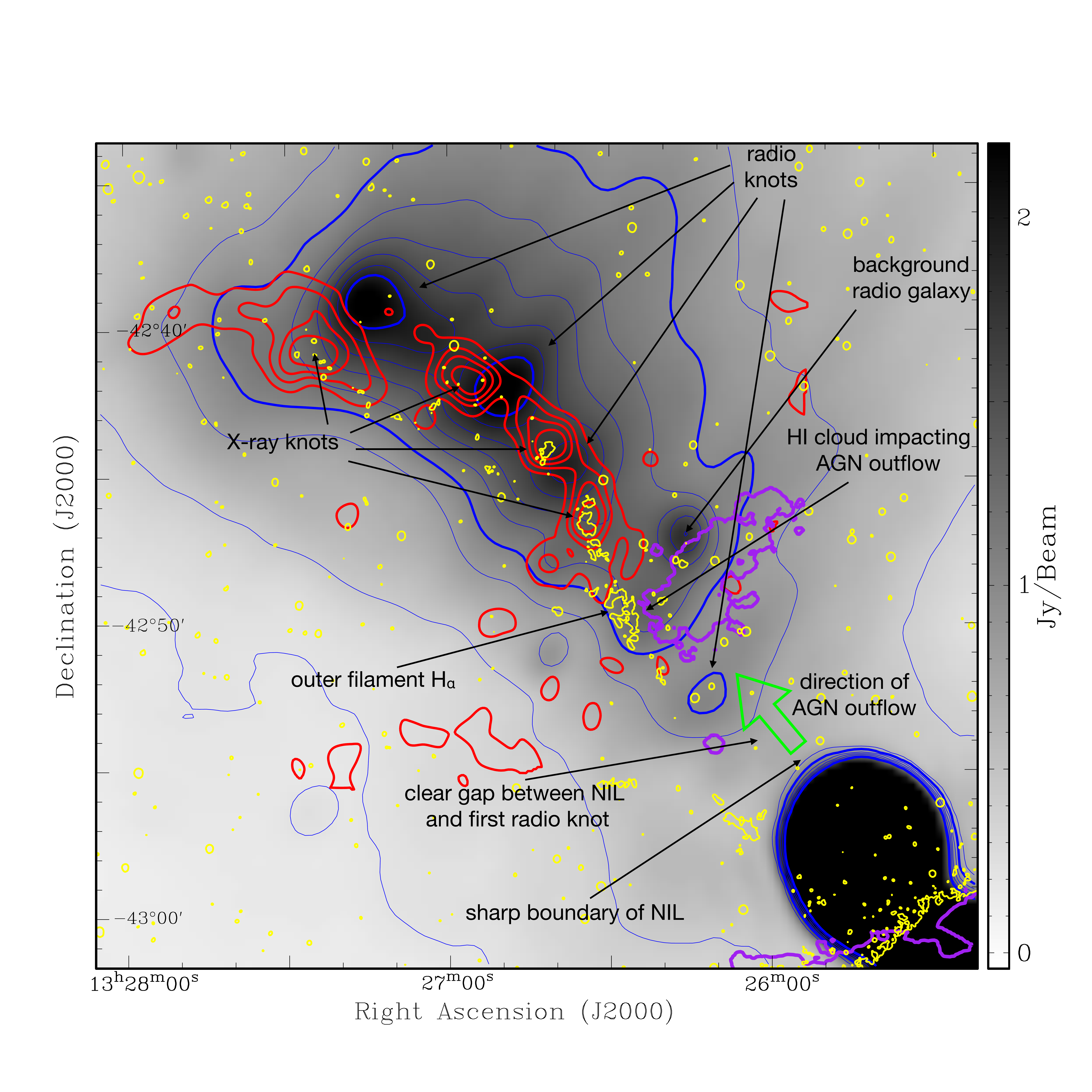}
\caption{The northern transition region with multifrequency data overlaid. Purple contour (at 5$\sigma$ above background): HI gas\cite{struve2010}, yellow contour (at 3$\sigma$ above background): H$\alpha$ emission (this work, see Methods), red contours (at 2,3,4 and 5$\sigma$ above background): X-ray data\cite{kraft2009}, blue contours (from 0.2 to 2.0, incrementing by 0.2 Jy/beam) and greyscale: MWA 185~MHz image (see Fig.~\ref{fig:1}). Black regions are where the bright radio emission saturates the color scale. The average restoring beam of the MWA image is a Gaussian of width 1.5$\times$1.2~arcmin with a major axis position angle of 155\degr}
\label{fig:2}
\end{figure}

The northern transition region (see Fig.~\ref{fig:1}, top right panel and Fig.~\ref{fig:2}), is rich in features across the electromagnetic spectrum, including bright radio emission (commonly referred to as the Northern Middle Lobe; NML\cite{neff2015a,morganti1999}), X-ray knots\cite{kraft2009}, filaments of H-alpha and FUV\cite{neff2015b} emission, and clouds of HI\cite{struve2010} gas, as well as molecular gas and dust (see \cite{neff2015b} and references therein). These features indicate an environment with complex and ongoing interactions between gas of multiple phases and velocities, as predicted by the unified CCA model of AGN feedback and as seen as an emergent property of e.g. the high-resolution TNG50 cosmological simulation\cite{nelson2019}. Our radio image reveals details in this region (Fig.~\ref{fig:1}, top right panel), not seen previously due to lack of sensitivity and/or imaging artefacts\cite{feain2011,neff2015a,morganti1999}. A series of bright radio knots arc across a path toward the north-east and are flattened along a north-west axis (this flattening was not apparent in the previous observations showing the knots\cite{neff2015a}). Appearing to stream north-west from the flattened radio knots are faint, arcing filaments that follow the magnetic field direction\cite{junkes1993} in the region.


The leading scenarios for the origin of the X-ray and radio knots are that they are due to heated or multiphase gas clouds, or that they are star formation sites\cite{neff2015a,neff2015b}. The flattened nature of the radio knots strengthen the argument for a gas-cloud origin. In this scenario, the multi-wavelength features of the NML (see Fig.~\ref{fig:2}) are being powered by a broad outflow from the AGN. The X-ray knots\cite{kraft2009} (red contours in Fig.~\ref{fig:2}) form where the mildly supersonic outflow impacts cold gas clouds in the region. The gas is compressed, bow shocks form upstream, and the clouds' outer layers are heated to X-ray temperatures and ablated. This is seen in simulations of starburst winds impacting clouds of cold gas\cite{cooper2009,gronke2018}. Radio knots, anti-coincident with the X-ray knots, form in front of the heated, X-ray-emitting clouds, either as electrons encounter compressed magnetic fields and/or are accelerated at the bow shocks. An available source of gas to form the clouds is a partial ring of HI (Fig.~\ref{fig:2}, purple contour\cite{struve2010}), which is either the result of CCA processes or tidal disruption events. See Methods for a detailed discussion of the kinematics, energetics and timescales of these northern transition region gas clouds. 

The arcing radio filaments replicate the shape and orientation of arcs of ionised gas observed on much smaller scales (<1~kpc) in the outer\cite{santoro2015b} and inner\cite{hamer2015,crockett2012} optical filaments. The arcing optical structures (see Extended Data Figure~\ref{fig:4}) bear a striking resemblance to features produced in simulations of broad transonic/supersonic winds\cite{gronke2018}, or expanding jet cocoons\cite{crockett2012}, impacting cold gas clouds. The matching shape and orientation of these optical arcs, and the radio filaments and flattened knots, indicates that they are being produced by the same outflow, which is much broader and far-reaching than what can be gleaned from observations of the optical filaments alone. Previous authors\cite{hamer2015} have suggested that the orientation of the inner filament arcs and velocity structure of the gas could be explained by the clouds being part of a backflow toward the AGN. Such inflowing gas is predicted by the CCA model of self-regulated AGN feedback\cite{gaspari2020,gaspari2017} and could exist alongside the outflow.

Another important radio feature in Fig.~\ref{fig:2} is the sharp edge of the northern inner lobe (NIL) and a clear gap between it and the first flattened radio knot. This strongly disfavours the existence of a previously-claimed\cite{morganti1999}, highly-collimated, `large-scale jet' of relativistic material escaping from the northern inner lobe, confirming the interpretation of images of the NML at 327~MHz\cite{neff2015a,neff2015b}, that there is no tightly collimated supply of relativistic particles from the northern inner lobe to the NML. 




Since we expect outflows from the AGN to be bipolar\cite{nelson2019}, it is interesting to note the stark difference between the radio morphology of the northern and southern transition regions. As shown in Fig.~\ref{fig:1}, bottom-right panel, the region to the south of the inner lobes, down to a declination of around $-44\degree$, displays none of the features seen to the north, excepting a single semi-circular filament that lies just to the east of the bright, extended background radio galaxy MRC1318-434B. Bordering this region, we highlight the `southern radio arc'. This radio arc stands out for two main reasons; unlike other similar features in the image it is roughly concentric with the active nucleus and, it borders an area otherwise devoid of radio features (particularly if we consider that the smaller, closer arc may be in the foreground, appearing closer to the AGN only in projection). Given these distinctions, one plausible explanation for the southern radio arc is that it represents the zone where a broad transonic outflow, as identified to the north, transitions toward a strong interaction with the ambient medium in the south. The contrast in morphology to the north and south can be explained by differing environments, a scenario that is strengthened by considering the merger history of Centaurus~A, which is traced by the presence of old, Red Giant Branch (RGB) stars.

\begin{figure}
\includegraphics[clip,trim=90 0 50 0,width=0.5\textwidth]{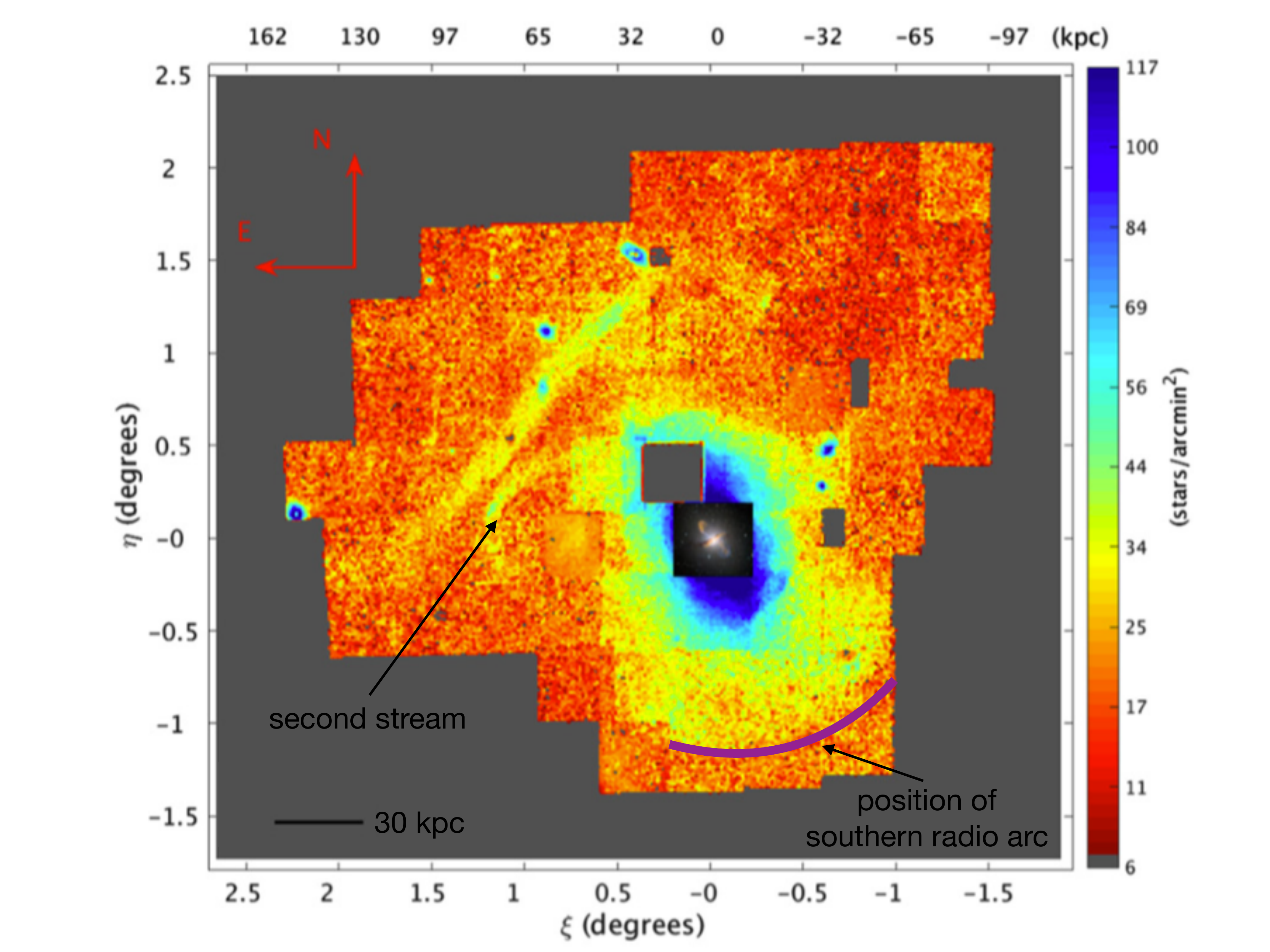}
\caption{Density map of RGB stars from Crnojevi\'c et al., 2016\cite{crnojevic2016} with the position of the $\sim90$~kpc-long southern radio arc from Fig.~\ref{fig:1} shown as a magenta line at a radius of approximately 75~kpc.}
\label{fig:3}
\end{figure}

In Fig.~\ref{fig:3} we reproduce a stellar density map of RGB stars\cite{crnojevic2016} and mark the location of the $\sim$90-kpc-long southern radio arc. There is a remarkable correspondence between the arc and the southern edge of a broad extension of RGB stars, which is not mirrored to the north of the galaxy. This stellar morphology can be largely accounted for by a major merger $\sim$2~Gyr ago, as shown in simulated stellar density maps of Centaurus~A\cite{wang2020}. The additional tidal streams present in the actual data\cite{crnojevic2016}, require infalling sub-structures at later times and in the northern region only. The ‘second stream’ identified by \cite{crnojevic2016} (see Fig.~\ref{fig:3}) is likely the remnant of a tidally disrupted dwarf, which has left behind dense gas in the northern transition region that may have disrupted the broad outflow in the north, while leaving the outflow unimpeded to the south. Such scenarios are seen in simulations\cite{nelson2019}, where both hierarchical structure formation and subsequent tidal disruption events produce asymmetric gaseous outflows similar to those observed in Centaurus~A (e.g. see fig.~2 of \cite{nelson2019}). 

With the above considerations, we contend that the southern radio arc approximately corresponds to the thermalisation radius, defined as the radius where the pressure from a large-scale outflow equalises with the hot halo pressure\cite{gaspari2017}. Future studies, including a wider radio frequency range and X-ray data, are planned to test this interpretation. From Fig.~\ref{fig:1}, we estimate the thermalisation radius as 75~kpc and the opening solid angle of the outflow as 1.2~sr. We assume that the outflow is bipolar and that we do not see a similar, clear radio feature correspondingly in the north, due to the gas asymmetries discussed earlier. Using these values and modified (to include the effect of a limited opening angle) versions of the equations describing the unified CCA model of AGN feedback\cite{gaspari2017}, we calculate the expected large-scale X-ray halo temperature and deduce a number of other AGN feedback properties for Centaurus~A (see the Methods section for testing CCA, the full details of the related calculations, and the comparisons with current Centaurus~A observations via multiple scales and wavelengths). The inferred X-ray halo temperature is 0.55~keV, giving a large-scale outflow speed of $\sim$1100~km s$^{-1}$ and a large-scale mass outflow rate of $\sim$2.9 $M_{\odot}$ yr$^{-1}$.

Having established the properties of the macro-scale outflow, we then calculate the corresponding feedback and feeding properties at small scales (Table~\ref{table:1}). We infer a small-scale outflow speed of $\sim$7100~km s$^{-1}$, a mass outflow rate of $\sim$0.07 $M_{\odot}$ yr$^{-1}$ and a mass \emph{accretion} rate onto the black hole (horizon) of $5.3\times10^{-4} M_{\odot}$ yr$^{-1}$. We refer to Methods for the full details of these calculations and a comprehensive comparison with current Centaurus~A observations, covering an unprecedented range to link the macro, meso, and micro scales (as suggested by the unification diagram in \cite{gaspari2020}).

The feedback properties we derive are compatible with current Centaurus~A data, consistent with relevant cosmological simulations and compatible with observations of AGN outflows at higher redshifts (see Methods). Hence, we demonstrate that Centaurus~A's proximity can be used to observe AGN outflows and inflows across ten orders of magnitude in physical scale and across a wide range of wavelengths. With the aid of high-resolution cosmological simulations, the insights gained from Centaurus~A may be used in future work to better understand galaxies at higher redshifts. Overall, Centaurus~A provides a key link between theory and observations by jointly probing multiple scales and phases, which will fundamentally advance our understanding of feedback and feeding throughout the local and distant Universe.
\newpage

\textbf{METHODS} \footnote{
$^{1}$International Centre for Radio Astronomy Research, Curtin University, Bentley, WA 6102, Australia
$^{2}$INAF, Osservatorio di Astrofisica e Scienza dello Spazio, via P. Gobetti 93/3, 40129 Bologna, Italy
$^{3}$Department of Astrophysical Sciences, Princeton University, 4 Ivy Lane, Princeton, NJ 08544, USA
$^{4}$Harvard-Smithsonian Center for Astrophysics, 60 Garden Street, MS-4, Cambridge, MA 02138, USA
$^{5}$Department of Geology \& Geophysics, Louisiana State University, Baton Rouge, LA 70803, USA
$^{6}$Netherlands Institute for Radio Astronomy (ASTRON), Postbus 2, 7990 AA Dwingeloo, The Netherlands
$^{7}$Kavli Institute for Astrophysics and Space Research, MIT, Cambridge, MA 02139, USA.
$^{8}$Department of Astronomy, Joint Space-Science Institute, University of Maryland, College Park, MD 20742, USA 
$^{9}$School of Natural Sciences, Private Bag 37, University of Tasmania, Hobart, TAS 7001, Australia 
$^{10}$ARC Centre of Excellence for All Sky Astrophysics in 3 Dimensions (ASTRO-3D), Bentley, WA 6102, Australia
$^{11}$Canadian Astronomy Data Centre, 5071 West Saanich Rd, Victoria, British Columbia V9E 2E7, Canada
$^{12}$Sydney Institute for Astronomy, University of Sydney, NSW 2006, Australia 
$^{13}$University of Tampa, 401 West Kennedy Boulevard, Tampa, FL 33606, USA 
$^{14}$Dunlap Institute for Astronomy and Astrophysics, University of Toronto, 50 St. George St, Toronto, ON M5S 3H4, Canada
$^{15}$Curtin Institute for Computation, Curtin University, GPO Box U1987, Perth, 6845, WA, Australia
}\\ 

\textbf{Radio observations and data processing}\\

The MWA\cite{tingay2013,bowman2013} was used to produce the radio image shown in Fig.~\ref{fig:1}. The MWA is a low-frequency interferometer located at the radio-quiet future site of the low frequency Square Kilometre Array\cite{skalow} in Western Australia. It has recently undergone an upgrade to phase 2, which has seen an increase in the maximum baseline length from 3 to 5.3~km\cite{wayth2018,beardsley2019}. Our image uses approximately 4~hrs of data split evenly between phase-1 (with better low-surface-brightness sensitivity) and phase 2 (with better angular resolution), using the full MWA instantaneous bandwidth of 30.72~MHz and a centre frequency of 185~MHz. Full details of the observations are given in Table~\ref{table:2}. 
Particular care was taken to ensure the quality of data used. Radio frequency interference (RFI) flagging was performed on the full-resolution data using \sc{aoflagger}\rm\cite{aoflagger1,aoflagger2} during the standard pre-processing stages of the MWA pipeline and each measurement set was manually inspected using tools such as \sc{aoquality }\rm and \sc{aoqplot}\rm\cite{aoflagger2}. Bad antennas and baselines were manually flagged and this inspection and flagging was repeated post-calibration. 

Due to the complexity of Centaurus A and the lack of a detailed and accurate source model at low frequencies, we performed initial calibration using dedicated calibration scans of the bright radio source Pictor~A. Phase and amplitude calibration were performed using \sc{calibrate}\rm\cite{calibrate} on a per-fine-channel basis, individually for each observation. Calibration solutions from the corresponding day (the phase-stability of the MWA allows us to use calibration solutions separated from target observations by several hours) were applied to target observations of Centaurus A. An initial set of 12 minutes of observation from both phase 1 and phase 2 (24 minutes in total) was then jointly imaged and deconvolved with \sc{wsclean}\rm\cite{wsclean}, using an implementation of the image domain gridding (IDG\cite{idg}) algorithm. The implementation of IDG, along with integration of the MWA primary beam pattern\cite{mwabeam}, allows accurate joint deconvolution of multiple MWA observations with different pointing directions (and hence different primary beam shapes). The limit of 24~minutes of observation was due to memory and CPU time considerations. 

The resulting image from 24~minutes of observation was limited in quality in the outer regions, due to radial artefacts emanating from the bright inner lobes. However, the model produced by \sc{wsclean }\rm of the inner region, including the inner lobes and the NML, was of sufficient fidelity to be used as the calibration model for the remaining observations. Hence, another seven images were produced from $\sim$24~minutes of observation each, with a direct source-model calibration and joint deconvolution. The image quality of each of these eight individual images varied, probably due to differing ionospheric conditions during the observations, with the main limiting factor being the radial artefacts similar to those seen in the initial image. Due to the different \emph{u,v} coverages of the data, the radial lines were attenuated when the eight images were averaged together. We used a simple weighting scheme based on the root-mean-square (rms) noise in a region to the south-west of the AGN to form a weighted average of the eight images, which is the final product shown in Fig.~\ref{fig:1}. Details including the Gaussian restoring beams and rms noise levels of the individual images, and the final image, are shown in Table~\ref{table:2}.

The final image has an angular resolution characterised by an average Gaussian restoring beam of width 1.5$\times$1.2~arcmin with a major axis position angle of 155\degr, a root-mean-square background noise level of approximately 4~mJy/beam and a peak brightness of 202~Jy/beam (giving a dynamic range of $\sim$50,000). It is the most detailed and accurate image of the whole radio source to date, particularly within 1$\degree$ of the inner lobes, where previous MWA\cite{mckinley2018} and ATCA/Parkes\cite{feain2011} images suffered from artefacts that obscured the details of the north and south transition regions (this region was also not covered in the spectacular polarisation images of the southern outer lobe from ASKAP\cite{anderson2018}, for the same reason). The radio galaxy spans approximately $8\degree$ on the sky along its north-south axis (the largest angular extent of any radio galaxy by a factor of~8), but is captured entirely within the wide field of view of the MWA, eliminating any need for mosaicing. Crucially, and unlike previous images of Centaurus A\cite{feain2011,mckinley2018}, the image is free from significant artefacts, with the exception of some faint remaining radial striping that does not affect our analysis. Within the inner $\sim1\degree$ radius of the AGN we achieve a factor of $\sim$500 improvement in dynamic range compared to the next best image of the entire source with a similar angular resolution\cite{feain2011}.  

Previous images covering just the NML region of Centaurus A have also suffered from significant imaging artefacts. The `large scale jet' identified in ATCA images\cite{morganti1999} at 1.4~GHz and used to justify many theories of the NML's origin over the last $\sim$20 years, is not present in our radio image, nor the relatively recent VLA observations of the NML at 327~MHz\cite{neff2015a}, indicating that it is in fact an imaging artefact. While the VLA observations failed to detect the large-scale-jet, they were also hampered by severe image distortions due to the low elevation of Centaurus A when viewed from the VLA site. 

In order to deal with the extreme dynamic range (which results in saturation of the central $\sim$1$\degree$ of Fig.~\ref{fig:1}, left panel) and to show all of the radio features in a single image, the intensity scale was manipulated to produce Extended Data Figure~\ref{fig:3a}. In the right panel of Extended Data Figure~\ref{fig:3a} the image has been stretched using the \sc{MaskedStretch }\rm tool of the astrophotography program \sc{PixInsight} \rm (https://pixinsight.com) and denoised to reduce the faint radial striping from the core. The \sc{MaskedStretch }\rm algorithm brings out all of the features present in the image over a very large dynamic range by iteratively performing a series of weak non-linear stretches, while masking the image with the result of the previous iteration at each step, so that high values are not clipped. A mean target background value of 0.05~Jy/beam was used and the program ran through 100 iterations. In the left panel of Extended Data Figure~\ref{fig:3a}, color has been added using \sc{Photoshop}\rm. Yellow and white indicate higher intensities, while red and magenta represent lower intensities. The post-processing image manipulations used in Extended Data Figure~\ref{fig:3a} produce a striking pair of images that display the key features of the radio galaxy, however their non-linear nature renders the images unsuitable for quantitative analyses. \\




\textbf{Optical observations and data processing}\\

The yellow H$\alpha$ contour in Fig.~\ref{fig:2} is from widefield optical data obtained using an astrophotography setup between 2019 June and 2020 February. The observations were made from R\'{i}o Hurtado, Chile, utilizing a Takahashi TOA-150 refracting telescope and Finger Lakes Instrumentation (FLI) ML~16200 monochrome CCD operated at $-25\degree$~C, onboard an Astro-Physics AP~1600 mount. The light was filtered using a set of luminance (L), red (R), green (G), blue (B), and 8~nm H$\alpha$ filters from FLI, with total exposure times of 13.25, 7, 7.5, 7.25, and 32.5 hours, respectively, giving a combined exposure time for the RGB image of 67.5~hours. The light frames were calibrated using 30 dark frames, 30 flat frames per filter, and 50 bias frames taken under identical conditions as the light frames, before being aligned and integrated using the winsorized-sigma rejection algorithm in \sc{PixInsight}\rm. Fig.~\ref{fig:2} displays a contour from the H$\alpha$ filter image that was continuum-subtracted utilising a technique adapted from Edoardo Luca Radice (http://www.arciereceleste.it/tutorial-pixinsight/cat-tutorial-eng/85-enhance-galaxy-ha-eng) whereby the median values of a significantly dimmed R frame were subtracted from the H$\alpha$ frame in order to remove residual red light outside of the H$\alpha$ band. 

Extended Data Figure~\ref{fig:4} shows a composite of optical images made using data from the Canada France Hawaii Telescope (CFHT), described below, and previously published\cite{mckinley2018} data from the Maryland-Magellan Tunable Filter (MMTF\cite{MMTF}) on the Magellan-Baade 6.5-meter telescope. The advantage of the CFHT is its wide field of view, which allowed us to image a square degree centred on NGC5128, and capture both the north and south transition regions, while the MMTF offers better resolution and higher image fidelity. The CFHT data were obtained using a combination of broad- and narrow-band filters with the Megacam instrument, via director's discretionary time (PI: Calzadilla; Proposal ID: 18AD88). Observations at H$\alpha$ (Ha.MP9603; $\Delta\lambda=104$\AA) were split into $5\times870$s exposures with a large dithering pattern (LDP1) and queued between 06-18 June 2018, resulting in a total exposure time of 4350s for this filter. An additional $5\times120$s exposures of archival $r$-band data (r.MP9601; $\Delta\lambda=1480$\AA) used in this paper were retrieved from the Canadian Astronomy Data Centre (Proposal ID: 03AE01). 

The data reduction was performed using the \texttt{ELIXIR} pipeline\cite{2004PASP..116..449M} to create bad pixel masks and correct for bias and flat fielding. Photometric calibration of the new and archival CFHT data was done with \texttt{MegaPipe}\cite{2008PASP..120..212G} using photometric standards that overlapped with Pan-STARRS coverage whenever possible, or using images taken during the same observing runs, to set up a zero-point. The H$\alpha$ photometric calibration is accurate to approximately 0.05 mag, and less for the archival $r$-band data. Astrometric calibrations are based on GAIA data and have errors of $\sim$30 mas. The images were all registered to the same pixel grid and co-added using \texttt{SWARP}\cite{2010ascl.soft10068B}. Background removal was performed using a local 128$\times$128 pixel mesh.

The large-scale CFHT image shown in Extended Data Figure~\ref{fig:4} is a continuum-subtracted H$\alpha$ map, meant to exclusively highlight the extended line emission throughout the entire square degree FOV. The continuum level in the H$\alpha$ 
map was estimated from the broader 
$r$-band filter and then subtracted after matching their point spread functions (PSFs). To model the PSF in both narrow and broadband images, we extracted 25$\times$25 pixel image cutouts centered on source catalogs produced via \texttt{SExtractor}\cite{1996A&AS..117..393B}, each normalized by the total flux in the cutout, and used the median of these normalized cutouts to create an ``average star''. The average star in both bands was re-centered and fit by a two-dimensional Moffat function to model the PSF with (H$\alpha$: ampl=0.034, $\gamma$=3.58, $\alpha$=2.37; $r$-band: ampl=0.017, $\gamma$=5.70, $\alpha$=2.68). We used the \texttt{photutils} (https://photutils.readthedocs.io/en/stable/psf.html) routine \texttt{create\_matching\_kernel} with a cosine bell window function to calculate the ratio of Fourier transforms needed to convolve the H$\alpha$ map with to match the broader PSF of the $r$-band image. The resulting convolved H$\alpha$ map was then subtracted by the $r$-band image, scaled by an arbitrary value to match the peak of the average star in both images, to produce a continuum-subtracted H$\alpha$ map. The asymmetric residuals present after the continuum subtraction can be attributed to distortions over the entire FOV due to the difficulty of achieving sub-pixel astrometric alignment over a square degree and 36 individual CCDs. In Extended Data Figure~\ref{fig:4}, these negative residuals were clipped and set to zero, and the image was stretched with asinh scaling to enhance the visualization of emission and suppress noise.\\




\textbf{Evidence for self-regulated feedback and feeding via Chaotic Cold Accretion}\\

There are a range of numerical models and zoom simulations, implementing AGN feedback in various ways\cite{turner2015,eisenreich2017,yoon2018,li2018,su2021}, which can reproduce many of the observed features of the galaxy population. The widely adopted paradigm in the literature is that there are two modes of accretion (excluding hierarchical mergers), that lead to black hole growth; the hot and the cold mode, based on whether the source of accretion fuel is hot or cold gas. Hot mode accretion predicts inflows and outflows that are mono-phase, continuous, and quiescent, whereas in the cold mode, accretion occurs in more intensive and intermittent bursts, at rates that can be orders of magnitude higher than the Bondi\cite{bondi1952} rate\cite{gaspari2019}. Large statistical samples\cite{best2012}, however, have shown that the underlying dichotomy is a result of the radiation efficiency of the AGN and that both hot and cold gas have roles to play in feeding and feedback\cite{hardcastle2018}. 

Hybrid models, which have the AGN cycling between the two modes of accretion, can explain observations of black hole masses and related galactic properties\cite{gaspari2016}. In these models, the steady and smooth flows described by Bondi accretion are disrupted by cooling instabilities, which result in Chaotic Cold Accretion (CCA\cite{gaspari2020,gaspari2013,gaspari2018,gaspari2017a}) of gas onto the black hole and an ensuing self-regulated feedback loop that prevents the establishment of catastrophic cooling flows\cite{gaspari2015}. The unified model of AGN feedback and feeding\cite{gaspari2017}, which incorporates CCA as the feeding mechanism, provides a number of clear observational predictions such as multiphase gas, turbulence, inflows and outflows, etc., which have been probed by several observations\cite{gaspari2020,gaspari2011a,gaspari2011b}, and also provides a framework that unifies the extreme range in spatial scale over which the feedback loop exists\cite{gaspari2013,gaspari2017}.



The observed properties of Centaurus~A on micro, meso and macro scales are well described by the unified CCA model of AGN feeding and feedback. As described in the main article and below in our discussion on consistency with other observations, the galaxy is surrounded by multiphase gas clouds with a wide range of temperatures, sizes and velocity dispersions. It is possible that some of this gas (such as the ring of HI at $\sim$10~kpc scales) is left over from previous tidal interactions with infalling galaxies, but CCA provides a robust mechanism for the replenishment of such clouds, which we show are destroyed on relatively short timescales by interactions with AGN outflows. There are also warm optical filaments and evidence for inflows and outflows on small, intermediate and large scales (see Table~\ref{table:1} and discussion below). VLBI observations indicate significant variability on the micro scales of the radio jet and the sharp boundaries of the inner lobes suggest that they are being inflated by a fresh outburst of activity not directly connected to the powering of the NML, implying that AGN activity is intermittent, with a duty cycle that is able to sustain a broad transonic outflow that pervades the meso and macro scales, causing weather systems in the NML and turbulent filaments in the outer lobes (the outflow may also be augmented by supernovae (SN) driven winds as discussed below in relation to simulations). These observational data, describing an intermittent AGN with multiphase inflows and outflows, clearly rule out a purely hot-mode accretion model and justify the application of the unified CCA model of AGN feeding and feedback to Centaurus~A throughout this paper.\\

\textbf{AGN feedback outflow/inflow calculations}\\

Our calculations are based on the equations of Gaspari et. al. (2017)\cite{gaspari2017}, hereafter G17, who use the CCA model of AGN feedback\cite{gaspari2013} to unify the ten-order-of-magnitude range in physical scale for energy transfer in active galaxies\cite{gaspari2020}, by leveraging a wide range of high-resolution simulations and multi-wavelength observations.
We interpret the long radio arc to the south of the Centaurus A core, shown in Fig.~\ref{fig:1}, as the thermalisation radius, $r_{\rm{th}}$, where the outflow pressure equalises with the ISM pressure. We measure $r_{\rm{th}}$ as approximately 75~kpc and estimate that the outflow has spread to a width of 90~kpc at a radius of 63~kpc, giving an opening angle of 71~degrees (as shown in Extended Data Figure~\ref{fig:3a}, right panel), resulting in a solid angle for the bipolar conical outflow of $\Omega\approx2.3$~sr. In this scenario we are viewing the edge of a segment of a sphere, which has its centre at the position of the AGN. As such, the estimate of $r_{\rm{th}}$ will not have a significant dependence on the inclination of the outer lobes (or the inner jets and lobes). In any case, our derived quantities are not overly sensitive to the exact value of $r_{\rm{th}}$ and uncertainties of up to 10\% do not affect our conclusions.

%
%
%

We use G17 equation~26 for the thermalisation radius, including the dependence on the bipolar outflow solid angle $\Omega$: 
\begin{equation} 
r_{\rm{th}} \simeq (55 \ {\rm kpc})\,\Omega_{4\pi}^{-1}\,T_{x,7.4} \simeq (295 \ {\rm kpc})\,T_{x,7.4}, 
\label{eqn:1}
\end{equation}
where $T_{x,7.4} \equiv T_x/10^{7.4}$~K is the X-ray temperature of the macro-scale hot halo (normalised to units of $10^{7.4}$~K $\simeq 2.16$~keV) and $\Omega_{4\pi}\equiv\Omega/4\pi$.

Rearranging Equation~\ref{eqn:1} to compute a value for the macro $T_x$ using our constrained $r_{\rm th} \simeq 75$~kpc, we obtain $T_x \simeq 0.55$~keV, which is similar to other X-ray constraints (see below). Following G17 equation 18, the micro-scale outflow velocity, $v_{\rm{out}}\simeq ({\rm 1.4\times10^4\,km\,s^{-1}})\,T_{\rm x,7.4}^{1/2}$, can be determined directly from the macro X-ray temperature, giving us $v_{\rm{out}} \simeq 7060$ km s$^{-1}$. 

We proceed by calculating the hot plasma entrainment factor, $\eta_{\rm{hot}}$, by also expanding G17 equation 23 to include the solid angle term $\Omega_{4\pi}$ and central ($r_0\approx1$\,kpc) density $\rho_{0,25}\equiv \rho_0/10^{-25}$ g cm$^{-3}$, such as:
\begin{equation} 
\eta_{\rm{hot}} = 40\ T^{-1}_{x,7.4}\,(\Omega_{4\pi}\,\rho_{0,25}\,r_{\rm{kpc}})^{2/3} = 9.4 \ T^{-1}_{x,7.4} \ r_{\rm{kpc}}^{2/3}, 
\label{eqn:2}
\end{equation}
where $r_{\rm{kpc}}$ is the radial distance from the nucleus (in kpc unit); in the last step, we have used the observed\cite{kraft2003} central hot plasma density of NGC\,5128, $\rho_{0,\rm{hot}}\simeq 6\times10^{-26}$ g cm$^{-3}$ (which is also close to the average assumed in G17) at a radius $r_{\rm{kpc}}=0.5$~kpc.
By using the retrieved macro $T_x\simeq0.55$~keV in Equation~\ref{eqn:2}, we obtain an entrainment factor $\eta_{\rm{hot}}\simeq42$. This is the loading factor by which mass is entrained by the micro outflow to produce the observed macro outflow. The macro outflow velocity $v_{\rm{OUT}}$ is related to the micro outflow velocity and the entrainment factor by G17 equation 21, $v_{\rm OUT} = v_{\rm out}/\eta^{1/2}$, providing $v_{\rm{OUT}}\simeq1090$ km s$^{-1}$.

Further, we use these values to compute the macro/micro mass outflow and inflow rates. The macro mass \emph{inflow} rate, also known as the (quenched) cooling rate, $\dot{M}_{\rm{cool}}$ is also tied to the macro X-ray temperature via the G17 equation 7, $\dot{M}_{\rm{cool}}\simeq ({\rm 1.1\,M_{\odot}\,yr^{-1}})\,T_{x,7.4}^2$, which gives us $\dot{M}_{\rm{cool}}=0.07\ {\rm M_{\odot}\,yr^{-1}}$. As discussed in G17, only a small percentage ($\sim$\,3\%) of the inflowing mass is actually accreted through the black hole horizon.  Most of the inflowing mass is expelled via the generated micro-scale outflow (converting binding energy into mechanical feedback energy), such that the micro mass outflow rate is $\dot{M}_{\rm{out}}\sim\dot{M}_{\rm{cool}}$. The macro mass outflow rate, $\dot{M}_{\rm{OUT}}$, is then given by the micro mass outflow rate multiplied by the entrainment factor (G17 equation 19); for our NGC\,5128 outflow; $\dot{M}_{\rm{OUT}}=\eta_{\rm{hot}}\dot{M}_{\rm{out}}\simeq2.9\,M_{\odot}$ yr$^{-1}$. To complete the loop and finally link the macro and micro scales, we calculate the micro mass inflow rate (i.e. the accretion rate onto the black hole horizon) by G17 equation 12 as $\dot{M}_{\bullet}=(3\%\,\dot M_{\rm cool})\,T_{x,7.4} \simeq 5.3\times10^{-4}$ $\rm M_{\odot}\,yr^{-1}$. These values are tabulated in Table~\ref{table:1}. See below for a discussion on the consistency of these calculations with observations at all wavelengths and physical scales.\\

\textbf{Consistency with other observations and theory}\\

In this work we have used low frequency radio observations to identify evidence for a broad, large-scale outflow from the AGN of Centaurus~A. Similar radio features to the southern radio arc, resulting from explosive outbursts at the centre of our own Galaxy, have been observed on smaller scales ($\sim400$~pc) at GHz frequencies\cite{heywood2019}. On more comparative scales (several kpc), the situation may be more analogous to the \emph{Fermi} bubbles\cite{su2010,dobler2010}, observed in $\gamma$-rays above and below the Galactic plane, with corresponding polarised GHz\cite{carretti2013} and microwave\cite{finkbeiner2004} radio emission. These bubbles can be explained by a weak but sustained outflow from the Galactic centre\cite{crocker2015}, where the expelled gas is unable to escape the Galaxy. In contrast, the AGN-driven Centaurus~A outflow is much more powerful, resulting in AGN feedback effects that extend ten orders of magnitude in physical scale, well beyond the optical extent of the galaxy. The recent discovery by the eRosita X-ray telescope\cite{erosita} of bubbles that are even larger and more energetic than the \emph{Fermi} bubbles in the Milky Way\cite{predehl2020}, indicates that our own galaxy may have once harboured an AGN, however the range of its influence would still have been markedly less than in the case of Centaurus~A.

We have used the G17 equations for the unified CCA model of AGN feedback to calculate the properties of the large-scale outflow, based on our radio image. From these large-scale (macro) properties, we have inferred the small-scale (micro) characteristics of the AGN outflow and inflow. Here we discuss the observational and energetic implications of these calculations and check for consistency with observations from the macro scale, through the intermediate (meso) scale, down to the micro scale (see Table~\ref{table:1}).

\textbf{Macro.} As discussed above, we derive a hot halo X-ray gas temperature of $T_x \simeq 0.55$~keV, for our measured thermalisation radius of $r_{\rm th} \simeq 75$~kpc. Current published X-ray observations of NGC~5128 lack the field of view required to image out to the thermalisation radius. However, using the much smaller fields of view of ROSAT (out to $\sim$35~kpc) and Chandra (out to $\sim$1~kpc), the hot halo temperature has been estimated at 0.3~keV\cite{kraft2009} and 0.6~keV\cite{bogdan2010}, respectively. Our value is within the bounds of these two estimates made with limited fields of view. Within the southern $\sim$1$\degree$/$\sim$60~kpc radius, which is free from clouds of cold gas, no features in the radio band or at other wavelengths (apart from the diffuse X-ray halo) are necessarily expected\cite{gaspari2011a,sadowski2017,gaspari2012}. With a larger field of view, forthcoming eRosita\cite{erosita} data may show X-ray features due to shocks at the thermalisation radius where we observe the southern radio arc. 

Using our derived outflow properties, we can derive some constraints on the energy budget of the system. We note here that the G17 model describes the final quasi-isotropic macro feedback deposition, mainly via wide entrained outflows and large cavities. In this model, narrow jet features can coexist, or even power the large scale outflows, but their micro effects are not considered. Our analysis therefore does not include the micro-scale effects of the Centaurus~A jets, but we do take into account that, on intermediate scales, previous jet outbursts have carved out a cocoon in the surrounding ISM/IGM, into which the outflow we detect is funneled. We first calculate the macro outflow power as:
\begin{equation} 
P_{\rm OUT}=\frac{1}{2}\dot{M}_{\rm{OUT}}v_{\rm OUT}^{2}\simeq1.1\times10^{42}\rm{erg \ s}^{-1},
\label{eqn:3}
\end{equation}
the pressure at the thermalisation radius, $p_{\rm th}$, is then given by:
\begin{equation} 
p_{\rm th}=\frac{P_{\rm OUT}}{v_{\rm OUT}A_{\rm th}}\simeq1.4\times10^{-13} \rm{dyn \ cm}^{-2}, 
\label{eqn:4}
\end{equation}
where $A_{\rm th}$ is the surface area of the spherical cap at radius $r_{\rm th}$, defined by the cone with solid angle $\Omega=1.15$~sr, as measured for the outflow to the south of the galaxy. We then calculate the energy of this bubble, $E_{\rm bubble}$, as:
\begin{equation} 
E_{\rm bubble}=4p_{\rm th}V_{\rm th}\simeq2.7\times10^{57} \rm{erg}, 
\label{eqn:5}
\end{equation}
where $V_{\rm th}$ is the volume of the spherical sector defined by $\Omega=1.15$~sr. This is the macro feedback energy being transmitted to the ISM/IGM (strictly speaking a lower limit, since it ignores shocks\cite{godfrey2016}). For our outflow with power $P_{\rm OUT}\simeq1.1\times10^{42}$~erg s$^{-1}$, the broad outflow would take $\sim$1.6 Gyr to deposit this amount of energy. Previous dynamical age estimates of the outer lobes, on the order of $\sim$1~Gyr\cite{eilek2014,wykes2013}, have been very sensitive to the inclination angle of the lobes to our line of sight. It is estimated from VLBI observations that the AGN jets are inclined at an angle $50<\theta_{\rm LOS}<80\degree$\cite{tingay1998}. If we assume that the southern radio arc represents the edge-on view of the compression zone, then our estimate of the thermalisation radius (and therefore the age of the outflow) is independent of the inclination angle of the jets. So, adopting the flow-driven model for the outer lobes\cite{eilek2014}, where the age is given by: $t_{\rm lobes}=600/\cos{\theta_{\rm LOS}}$~Myr, and using our age estimate for the macro outflow as the outer lobe age, then we arrive at an estimate for the inclination angle of $\theta_{\rm LOS}=68\degree$, in agreement with VLBI observations.

The thermal gas present in the outer lobes has been measured by studies of radio polarisation\cite{osullivan2013} and X-rays\cite{stawarz2013}. The number density of thermal gas derived from these observations is $\sim$10$^{-4}$cm$^{-3}$. If this thermal gas is due to entrainment of gas from the central galaxy, then the implied mass outflow rate is $\sim$20 $M_{\odot}\,$yr$^{-1}$, an order of magnitude greater than our inferred mass outflow rate (see Table~\ref{table:1}). It is possible, however, that these observational results are contaminated by Galactic foreground emission, a scenario that could not be ruled out by the authors\cite{osullivan2013,stawarz2013}, and has also been put forward as an explanation by other authors who estimate a much smaller thermal gas density of $\sim$10$^{-8}$cm$^{-3}$ for the outer lobes, based on entrainment by the currently active inner jets\cite{wykes2013}. Our mass outflow rate of $\sim$2.9 $M_{\odot}\,$yr$^{-1}$, based on a broad AGN outflow is therefore consistent with the scenario where the true thermal gas content of the outer lobes is on the order $\sim$10$^{-5}$cm$^{-3}$ and the remainder observed by radio polarisation and X-ray studies\cite{osullivan2013,stawarz2013} is due to Galactic foreground contamination.

\textbf{Meso. }We now turn our attention to the intermediate (meso) scales and in particular to the north of the galaxy, where the situation is vastly different to the south. This contrast is due to the presence of clouds of cold and warm gas in the path of the AGN outflow, as we have shown in Fig~\ref{fig:2}. The purple contour in Fig~\ref{fig:2} shows HI emission\cite{struve2010}, which is prominent in the disk of NGC~5128, but also exists in a ring-like structure surrounding the galaxy at a radius $\lesssim$17~kpc \cite{schiminovich1994}. The HI ring appears to be tilted on a similar angle to the radio jets and is moving such that the cloud to the north (near the centre of Fig.~\ref{fig:2}) is coming toward us with a line-of-sight velocity of around 400~km~s$^{-1}$ and the cloud to the south of the galaxy is moving away from us at around 700~km~s$^{-1}$\cite{schiminovich1994}. At the centre of Fig.~\ref{fig:2} the eastern edge of the northern HI cloud has entered into the outflow from the AGN, which has been funneled into the cavity created by previous outbursts of radio jet activity\cite{neff2015b}. As the cold gas impacts the outflow, an area of turbulence results (evident in the large random velocities observed in the filaments\cite{santoro2015b,graham1981,morganti1991,santoro2015a}), inducing star formation\cite{sutherland1993} and creating the distinctive ionised filament, which is bright in H$\alpha$ (Fig.~\ref{fig:2} yellow contour and Extended Data Figure~\ref{fig:4}) and Far-UV\cite{neff2015b} emission. 

While the differences in the environments to the north and south of the galaxy can be tied to its merger history, asymmetries in the warm and cold gas distributions around galaxies can occur in the absence of a merger event, as shown in simulations\cite{gaspari2012b}, resulting in one part of a bipolar outflow being hampered or deflected. At smaller radii in NGC 5128 ($\sim$3.5~kpc) there is evidence to suggest a sharp discontinuity in pressure due to non-hydrostatic gas motions or `sloshing' to the north of the galaxy\cite{kraft2008}, helping to explain asymmetries in the gas distribution and radio morphology of Centaurus~A.

The outflow parameters we derive are very similar to those of the `starbust wind' invoked by Neff et al. (2015)\cite{neff2015b}, hereafter N15, to explain the features of the NML. Above we calculated an outflow power of $P_{\rm OUT}\simeq1\times10^{42}$~erg s$^{-1}$, the same value used by N15 to estimate the power incident on the small gas cloudlets making up the inner and outer filaments. They find that this wind alone is insufficient to provide all of the necessary power\cite{morganti1991,sutherland1993} to sustain the emission-line filaments, but could provide enough power if boosted by a factor of 10 by the AGN jet flow. The current AGN jet power is estimated as $P_{\rm jet}\simeq2\times10^{43}$~erg s$^{-1}$\cite{neff2015a}, so it is not unreasonable to assume that similar powers have been available in previous AGN outbursts to boost the wind in such a manner, close to the jet axis. While the G17 equations, which focus on the macro feedback properties, do not describe such localised and narrow jet features, this scenario is still consistent with the CCA model and the broad outflow that we have identified. Given the close alignment of the emission-line filaments with the orientation of the inner lobes and jets, it is possible that the outflow could be sporadically boosted by a factor of 10 in power by a jet outburst, providing power to these short-lived (10-15~Myr\cite{neff2015b}) clouds, without the need for a `large -scale jet' currently connecting the inner lobes to the NML. However, more detailed simulations incorporating both broad winds and narrow jets impacting cold gas clouds are required to verify these claims.


The centre of Fig.~\ref{fig:2} also appears to be the nexus of the X-ray\cite{kraft2009} (red contours) and radio (this work, blue contours and greyscale) knots. The X-ray knots are analysed in detail by Kraft et al. (2009)\cite{kraft2009}, hereafter K09, who conclude that the most likely source of energy for the knots is the previously-discussed large-scale jet\cite{morganti1999}. Since we have dispelled the existence of this large-scale relativistic jet, we assess whether our transonic large-scale outflow can power the X-ray knots. As a minimum test of consistency, K09 assert that the ram pressure of their large scale jet must be greater than the pressure of the individual knots in order to contain them. Using the same values as K09 for the cloud cross-sectional area, the ram pressure for our outflow is $p_{\rm ram}=2P_{\rm OUT}/(v_{\rm OUT}A_{\rm clouds})\simeq5.9\times10^{-10} \rm{dyn \ cm}^{-2}$. This is $\sim$30 times the typical pressure of the K09 X-ray clouds, enough to prevent them from dissipating too quickly. The total thermal energy of the X-ray knots estimated by K09 is $2.6\times10^{55}$~erg. Given the outflow parameters that we calculate above, we estimate that the power intercepting the clouds is $1.4\times10^{40}$ erg s$^{-1}$, so it would take $\sim$56~Myr to provide the required energy to the clouds. This is an order of magnitude longer than the lifetimes of the clouds estimated by K09 of a few Myr. As discussed above for the smaller emission-line clouds of the filaments and below for the micro-scale duty cycle, the AGN jets can provide sporadic boosts to the outflow, which can make up the energy deficit, while the constant, broad outflow maintains the pressure required to contain the clouds.

Closer to the nucleus, but still in the meso-scale regime, other outflows of gas have been identified in Centaurus~A. Relatively dense and warm outflows $\sim$3~kpc to the east and west of the galaxy have been identified by a spectral analysis of Chandra X-ray data\cite{krol2020}, with an estimated outflow velocity of $\sim$1000~km~s$^{-1}$ and mass outflow rate of $\sim$1 $M_{\odot}\,$yr$^{-1}$. A slower, but potentially more massive outflow of cold neutral and ionised gas has also been detected within the inner 500~pc of Centaurus~A\cite{israel2017}, with a projected velocity of 60~km~s$^{-1}$ and mass outflow rate $\sim$2 $M_{\odot}\,$yr$^{-1}$ (uncertain by a factor of 3). The outflows probably result from eruptive events near the galactic nucleus that exist on smaller timescales (e.g. $\sim$3~Myr for the hot east-west outflows\cite{krol2020}) and don't necessarily have the same physical origin as our large-scale wind, but demonstrate that outflows with comparative mass outflow rates and velocities do exist in Centaurus A on these physical scales. It is interesting to note that there is currently little evidence in Centaurus~A for cold outflows on scales larger than 1~kpc\cite{veilleux2020,veilleux2005}. However, the currently available HI observations\cite{struve2010,schiminovich1994}, for example, do not cover out to the thermalisation radius that has been identified in this work. New instruments such as ASKAP\cite{askap1,askap2} and MeerKAT\cite{meerkat1,meerkat2} will soon probe these regions and may uncover so-far undiscovered cold gas clouds.

\textbf{Micro. }One interpretation of the sub-pc scale VLBI data on the Centaurus~A core\cite{tingay1998,tingay2001,jones1996} is that discrete jet components with subluminal motions of $\sim0.1c$ ($3\times10^{4}$ km s$^{-1}$) coexistent with a faster underlying jet flow with velocity $\sim0.45c$ ($13.5\times10^{4}$ km s$^{-1}$). X-ray observations have also been used to look for ultrafast outflows (UFOs) from the core of Centaurus~A\cite{tombesi2014}. Blueshifted Fe~K absorption lines indicate an outflow from Centaurus~A, variable with time, between 900 - 1500 km s$^{-1}$. These velocities are less than the $\sim$10$^{4}$ km s$^{-1}$ typical velocities of UFOs found in radio-loud AGN\cite{tombesi2014}, but this could be due to a different angular regime being probed because of the extreme proximity of Centaurus~A. In any case, our inferred micro-scale outflow velocity of $\sim$7060~km~s$^{-1}$ sits comfortably between these two observed micro outflow velocities. While we are not concerned in this work with the specific mechanisms that produce these outflows, we note here that AGN jets can produce such outflows\cite{nesvadba2016}, resulting in feedback at the macro scale, that occurs over a much broader cross section than the typical jet width\cite{morganti2015,bicknell2015}.

As for the inferred \emph{inflow} parameters, our results give a value of $\dot{M}_{\rm in}\sim$ 5.3$\times$10$^{-4}$ $\rm M_{\odot}\,yr^{-1}$ for the BH inflow rate, which is an average over the lifetime of the source and is related to the gas accreting through the BH horizon (hence the small value). 
The measured inflow rate of HI\cite{vangorkom1989,vanderhulst1983} at the micro scale is 5x10$^{-2}$ $\rm M_{\odot}\,yr^{-1}$ and since only $\sim$3\% of the actual inflowing gas is accreted through the black hole horizon\cite{gaspari2017}, these numbers are fully consistent. Interestingly, if we apply a dynamical jet model\cite{turner2015} to our inferred accretion rate and estimate the jet power as $P_{\rm{jet~model}}=\epsilon\dot{M}_{\bullet}c^{2}$, and using a reasonable value for the jet production efficiency, $\epsilon$, of 10\%\cite{turner2015}, we obtain $P_{\rm{jet}}\simeq3\times10^{42}$~erg~s$^{-1}$. This falls between our inferred outflow power of $P_{\rm OUT}\simeq1\times10^{42}$~erg s$^{-1}$, which is a measure of the average power over the lifetime of the outflow, and the estimated \emph{current} jet power $P_{\rm jet}\simeq2\times10^{43}$~erg s$^{-1}$\cite{neff2015a}. This indicates that our inferred values are reasonable and in agreement with a model\cite{turner2015} that invokes an AGN that intermittently switches on and off. A rough estimate for the duty cycle of the AGN activity can be obtained by calculating the fraction of `on' time needed to keep the hot halo in approximate heating-cooling balance. We estimate this by taking the ratio of the X-ray cooling luminosity (calculated from G17, equation 7 using our inferred value for $T_x \simeq 0.55$~keV, we obtain $L_x \simeq 1\times10^{41}$~erg~s$^{-1}$) and the jet power, $P_{\rm jet}\simeq2\times10^{43}$~erg s$^{-1}$, which gives us a value of 5\%. This is a reasonable value, which we note is well above the duty cycle that would be inferred from the mass-luminosity scaling relationship\cite{best2005} of 0.1\% (based on a black hole mass of 0.35 - 8.5$\times 10^{7}$\cite{neumayer2010} and a 1.4~GHz radio luminosity of 2.3$\times10^{24}$~W~Hz$^{-1}$\cite{cooper1965}), because this scaling relationship averages across all black holes of the same mass, most of which will have radio-quiet host galaxies, unlike Centaurus~A.

We can also calculate the (micro) mechanical efficiency as $\epsilon_{\rm BH}=P_{\rm OUT} /(\dot{M}_{\rm in}c^{2})=0.036$. We note that this value is consistent with the 3$\pm$1\% found via GR-rMHD simulations over a wide range of tested physics\cite{sadowski2017} and is far from the unity value needed to power relativistic jets via spin (e.g. \cite{mcnamara2009,mcdonald2021}). Indeed, such a large value would overheat and unbind most of the group halos.\\ \\



\textbf{Implications for cosmological simulations}\\

Feedback (through SN driven winds and black-hole accretion) affects galaxy evolution and cosmology, but is difficult both to simulate and to observe\cite{nelson2019,heckman2017,donnari2019}, which is due in no small part to the range of physical scales involved. Accretion onto black holes occurs on scales of $\sim$10$^{-5}$ pc, while the energy-driven outflows that this process launches imprint their signatures on scales of $\sim$10$^5$~pc, affecting the density, temperature, entropy, and cooling times into the far reaches of the gas halo\cite{zinger2020}. In fact, the effects of SMBH growth and associated feedback extend beyond the halo scale through cosmic gas flows and SMBH-SMBH mergers\cite{bassini2019,truong2021}, such that any simulation conducted in cosmic isolation will be incomplete. Hence, we turn our attention here to cosmological simulations and focus on how observations of Centaurus~A (presented here and planned for the future) can be used to inform simulations (and vice versa).

Broad, bipolar galactic scale outflows are a feature of modern cosmological simulations such as EAGLE\cite{mitchell2020}, SIMBA\cite{dave2019}, and Horizon-AGN\cite{beckmann2017}. In these simulations, both AGN and stellar feedback contribute to the outflows, and in the case of AGN-driven feedback, the directionality of the bipolar outflows along the minor axis of the galaxies are directly specified as part of the sub-grid physics implemented. Higher-resolution zoom simulations such as ROMULUSC\cite{tremmel2019} have been able to reproduce collimated large-scale outflows from isotropic AGN feedback introduced on small scales, showing that this emergent effect is a result of the morphology and angular momentum of the gas at the base of the outflow and not due to jet orientation. Such AGN feedback effects are difficult to implement in a full-scale cosmological simulation, as limited resolution affects the way that AGN outflows couple to nearby gas particles\cite{tremmel2019}.

A current set of key cosmological simulations that include AGN feedback is the IllustrisTNG suite \cite{marinacci2018,naiman2018,nelson2018,pillepich2018,springel2018}. The simulation runs TNG100 and TNG300 produced realisations of very large volumes ($\sim$100$^{3}$ and $\sim$300$^{3}$ comoving Mpc$^{3}$, respectively), allowing for statistical studies of large galaxy populations, at the cost of resolution to study smaller details of individual galaxies. The simulation run TNG50 \cite{nelson2019,pillepich2019}, simulated a smaller volume at a higher resolution, making it ideal to study feeding and feedback processes. The resolution of TNG50 is equivalent to or better than many `zoom' simulations that have previously identified fast galactic outflows of $\sim$1000~km s$^{-1}$ out to radii of 100s~kpc\cite{brennan2018}, but also allows the study of feedback effects on cosmological scales. 

An important result from TNG50 is that large-scale outflows are bipolar, with wide opening angles (e.g. around 70$\degree$), directed along the minor axis of the galaxy\cite{nelson2019}. This is a purely emergent property of the outflows, since the feedback inputs inserted at small scales (associated with both SN-driven winds and black-hole accretion) have no direction dependence. Our Centaurus~A observations are consistent with this scenario, i.e. despite the radio galaxy being characterised by highly collimated, narrow jets on sub-pc scales, the end result is a broad bipolar outflow affecting a wide region along the minor axis of the galaxy. This also corroborates the assumption of the CCA model, that broad, large scale outflows dominate over the effects of narrow jets at scales of 10s to 100s kpc.

Another important result from TNG50 is that outflows are multiphase, consisting of cold, warm and hot gas. Observational examples where multiple phases are observed in the one source, however, are rare\cite{nelson2019}, hence the importance of Centaurus~A, which provides a rich data set of multiphase gas observations. Of critical importance to the CCA model and cosmological simulations are the X-ray scaling relations, which allow many properties to be inferred from the hot gas halos of galaxies\cite{gaspari2019}. As discussed previously, X-ray observations of the hot gas halo of Centaurus~A have been limited by instrumental fields of view. Our radio observations, however, have allowed us to infer the hot halo temperature via a measurement of the thermalisation radius. While outside of the scope of this work, subsequent studies will therefore be able to compare the observed large and small scale, multiphase gas properties of Centaurus~A with analogs of the galaxy found in simulations such as TNG50.  \\

\textbf{Compatibility with observations at higher redshift}\\

The general consistency of the outflow properties of Centaurus~A with galaxies throughout the IllustrisTNG simulations gives us some confidence that Centaurus~A is not atypical in terms of its feedback processes. This is also supported by comparisons to galactic outflows observed in galaxies at higher redshifts. It is clear that outflows, particularly driven by stellar winds and supernovae, are ubiquitous in galaxies throughout the universe and across cosmic time\cite{veilleux2020}. These stellar processes must contribute to the observed outflow in Centaurus~A, which is currently undergoing a starburst\cite{neff2015b}, but energy from the AGN is required in order to reproduce the observed features in the transition regions. While it is not possible to separate the contribution of AGN activity and stellar activity in either simulations or observations\cite{nelson2019}, there is now a lot of evidence for outflows in AGN, obtained mainly through optical absorption and emission lines and X-ray observations. 

In the local Universe ($z\simeq0$ to $z\simeq0.2$), ionised outflows with velocities ranging from 1000~km~s$^{-1}$ to $0.4 c$ are observed in up to 50\% radio loud AGN\cite{tombesi2014,gofford2013,gofford2015}. These identifications via measurement of Fe~K absorption lines in the X-ray band favour the discovery of very fast outflows close to the nucleus, so slower outflows further out may be missed by this method. These ultrafast outflows on the order 10$^{4}$ km~s$^{-1}$ are consistent with our inferred micro outflow rate for Centaurus~A of 7060 km~s$^{-1}$. These observations demonstrate that the presence of AGN jets does not preclude the existence of broad outflows or winds, as we observe in Centaurus~A. This has also been shown by the identification of a $\simeq$1000~km~s$^{-1}$ outflow in the powerful local radio galaxy 3C~120\cite{tombesi2017}. 

At higher redshifts ($z > 0.6$), optical emission lines provide a means to identify outflows. Velocity widths of $>600$ km s$^{-1}$ indicate fast ionised outflows in 50 - 70\% of AGN and are more prevalent in radio-loud AGN\cite{harrison2016}. Around the peak of quasar space density at $z\simeq2$\cite{schmidt1995}, outflows are also found to be common in galaxies harbouring AGN\cite{leung2017,circosta2018,kakkad2020}. These outflows, also detected by means of optical emission lines are found to have velocities from 300 - 3000~km~s$^{-1}$. Gravitational lensing has allowed glimpses of ionised outflows with similar properties out to $z=3.91$\cite{gofford2013,gofford2015}, and even out to $z\simeq5$ there is evidence for molecular outflows in quasars, which appear to be dominated by AGN activity, rather than stellar processes\cite{bischetti2019}. 

It is clear that outflows, such as those observed in Centaurus~A, are a common and important feature of galaxies across cosmic time. However, their detection and characterisation is difficult, usually relying on a single tracer and incorporating many assumptions about geometry and galactic properties\cite{veilleux2020,nelson2019}. While we should be careful comparing Centaurus~A directly with high redshift galaxies, it does bear significant similarities to radio galaxies at high redshifts (e.g. residing in a group and being recently post-merger, unlike most in the local Universe). The availability of a multitude of different tracers and the ability to observe both micro and macro-physics in the one source, provide our best opportunity to study physical processes that are extremely difficult for more distant sources, such as outflow launching mechanisms and multiphase gas interactions, and to incorporate these processes into simulations in a coherent and self-consistent manner.\\

\textbf{Future developments}\\

We are still at the beginning of our journey to understand the details of AGN feeding and feedback and implementing them into cosmological simulations in a complete sense. Centaurus~A presents a unique opportunity for us to use a multi-wavelength, multiscale approach, combining theory and observations, to thoroughly understand the multiphase inflows and outflows that are ubiquitous across the Universe. As observations move toward higher resolution and sensitivity and the challenges of widefield imaging are addressed, our data on Centaurus~A will continue to improve, allowing us to fill in the many gaps in our Table~\ref{table:1}. Detailed and accurate simulations spanning at least ten orders of magnitude (and incorporation of models into cosmological simulations), are required if we are to fully make sense of these data. How is energy transported across the various scales and what are the duty cycles for feedback, how does this correlate to galaxy evolution over Cosmic time? These questions and many more\cite{gaspari2020} can be tackled using the unique resource that is our nearest radio galaxy.\\



\begin{table*}
	\centering
	\caption{Unified AGN Feedback Properties of Centaurus~A}
	\label{table:1}
	\begin{tabular}{lccccccr} 
		\hline
		Description$^{\rm{a}}$ & Radius/ & Temp.$^{\rm{c}}$ &  Num. Density$^{\rm{d}}$ & $v_{\rm{out}}^{\rm{e}}$ & $\dot{M}$$^{\rm{f}}$ & $\sigma_{v}$$^{\rm{g}}$ & Reference(s)$^{\rm{h}}$\\
		  & Scale$^{\rm{b}}$ (kpc) & ($10^{6}$ K) & (cm $^{-3}$) & (km s$^{-1}$) & ($M_{\odot}$ yr$^{-1}$) & (km s$^{-1}$) & \\
		\hline
		\hline
		\textbf{Micro (mpc-pc)} &  &  &  &  &  &  &\\
		Inferred inflow (horizon) & <1 x $10^{-6}$ & - & - & - & 5.3 x $10^{-4}$ & - & This work\\
		Inferred outflow & <1 x $10^{-4}$ & - & - & 7060 & 0.07 & - & This work\\
        Radio jet outflow & 1 x $10^{-3}$ & - & - & (3 - 13.5) x $10^{4}$ & - & - & \cite{tingay1998,tingay2001}\\
		Fe~K absorption outflow & <2 x $10^{-3}$ & 7.5 & - & 900 - 1500 & - & 1000-5000 & \cite{tombesi2014}\\
		\hline
        \textbf{Meso (pc-kpc)} &  &  &  &  &  &  &\\
        CND North ionised outflow & <0.5 & 0.035 & 10 (e-) & 60 - 95 & 0.6 - 5.4 & <200 & \cite{israel2017}\\
        CND North neutral outflow & <0.5 & .00016 & 4 x $10^{3}$ & 60 - 95 & 0.6 - 5.4 & <200 & \cite{israel2017}\\
		East/West outflow & 2.7 & 2.3 (0.2 keV) & 0.3 - 0.5 & 600 - 1000 & 0.5 - 1.7 & - & \cite{krol2020}\\
		South inner lobe expansion & $\sim$5 & 11 & 0.004 & 2600 & - & - & \cite{croston2009}\\
		North inner lobe expansion & $\sim$5 & 0.16 & - & - & - & - & \cite{krol2020}\\
        N Inner ionised filament & 8 & - & - & - & - & 50 - 140 & \cite{neff2015b,hamer2015,morganti1991}\\
        N Outer ionised filament & 15  & - & - & - & - & 50 - 140 & \cite{neff2015b,santoro2015b,morganti1991,santoro2015a}\\
        S ionised cloud & 12 & - & - & - & - & - & \cite{keel2019}\\
        HI ring & <17 & cold & - & - & - & 80 & \cite{struve2010,schiminovich1994}\\
		NML X-ray knots (N1-4) & 15-30 & 5.8 - 12 & .008 - .012 & - & - & - & \cite{kraft2009}\\
		\hline
        \textbf{Macro (kpc-Mpc)} &  &  &  &  &  &  &\\
        Hot X-ray halo & $\geq$35  & 0.3-0.6~keV & 0.03-0.01 (e-) & - & - & - & \cite{kraft2009,kraft2003,bogdan2010,osullivan2003}\\
        Large-scale outflow & <75  & 0.55 keV & - & 1090 & 2.9 & - & This work\\
        Inferred cooling inflow (quenched) & <75 & - & - & - & 0.07 & - & This work\\
		Thermal gas in lobes & <800 & 5.8* &  $10^{-4}$* & - & 20 - 200* & - & \cite{osullivan2013,stawarz2013}\\
		\hline
		\multicolumn{8}{|c|}{Notes: (a) Relevant AGN feedback feature e.g. inflow or outflow, gas cloud/knot/filament (b) Approximate radius from the nucleus (c) Temperature of the gas} \\
		\multicolumn{8}{|c|}{in units of $10^{6}$K or in keV when more relevant (d) Number density of gas, this is proton density unless labelled (e$^{-}$) for electron density (e) Outflow velocity} \\ 
		\multicolumn{8}{|c|}{(f) Mass outflow rate (or inflow rate if labelled (in) (g) Velocity dispersion (h) References *These values may be excessively high due to Galactic foreground} \\
		\multicolumn{8}{|c|}{contamination, see discussion in Methods.}\\
		\hline
	\end{tabular}
	
\end{table*}

\begin{table*}
	\centering
	\caption{Details of MWA radio observations}
	\label{table:2}
	\begin{tabular}{lcccccccr} 
		\hline
		Image & Phase 1 obsids & Phase 1 dates & Phase 2 obsids & Phase 2 dates & Int. time & bmaj,bmin,bpa & rms noise\\
		group &  &  &  &  & (min) & (arcmin,arcmin,deg) & (mJy/beam)\\
		\hline
		\hline
		Cen~A & 1117031728, 1121334536  & May-July  & 1199663088, 1200604688  & Jan  & 33.1 & 1.37, 1.12, 155.0  & 10.8 \\
		  Im. 1      & 1121507392, 1121593824  & 2015 &                1200777584, 1200864032  & 2018 &    & & \\
		        & 1121420968, 1121680256  &  &                1200691136, 1200950480  &  &   & & \\
		\hline
		Cen~A & 1122198832, 1122112400  & July  & 1201123376, 1201296272  & February  & 35.2  & 1.48, 1.23, 163.0 & 15.9 \\
		  Im. 2      & 1121939544, 1121853112  & 2015 &                1201469160, 1201555608  & 2018 &   & & \\
		        & 1122025976, 1121766680  &  &                1201382720, 1201814952  &  &   & & \\
		\hline
		Cen~A & 1122285264, 1122371696  & July-August  & 1201895248, 1201900584  & February  & 31.5  & 1.51, 1.29, 163.4 & 12.3\\
		  Im. 3      & 1122544552, 1122630984  & 2015 &                1201981408, 1201986752  & 2018 &   & & \\
              & 1122458120, 1122717416  &  &                1201901392, 1201987840  &  &   & & \\
        \hline
		Cen~A & 1112806040, 1112892200  & April-May  & 1202239904, 1202326064  & February  & 23.2  & 1.40, 1.06, 145.1 & 9.50 \\
		  Im. 4      & 1114869144, 1114955312  & 2015 &                1202411608, 1202418952  & 2018 &   & & \\
		       & 1114782984, 1115041472  &  &                1202410528, 1202672864 &  &   & & \\
        \hline
		Cen~A & 1115049272, 1115127640  & May  & 1202673200, 1202673440  & February  & 23.2  & 1.44, 1.15, 158.9 & 9.85 \\
		  Im. 5      & 1115213800, 1115221600  & 2015 &   1202673920, 1202674160  & 2018 &   & & \\
		        & 1115135440, 1115299968  &  &   1202673680, 1202678472  &  &   & & \\
		\hline
		Cen~A & 1115307768, 1116343632  & May  & 1202679384, 1202756888  & February  & 24.1  & 1.37, 1.16, 159.9 & 17.4\\
		  Im. 6      & 1116603776, 1116604056  & 2015 & 1202843048, 1202851152  & 2018 &   & & \\
		       & 1116429792, 1116773232  &  & 1202764984, 1203015384  &  &   & & \\
		\hline 
		Cen~A & 1120470256, 1120815968  & July  & 1242739136, 1243169056  & May-June  & 28.8  & 1.58, 1.21, 147.3 & 9.28\\
		  Im. 7      & 1120988824, 1121075248  & 2015 &                1243859272, 1244376256  & 2019 &   & & \\
		        & 1120902392, 1121161680  &  &                1243341384, 1244807072  &  &   & & \\
        \hline
		Cen~A & 1121248104, 1121334536  & July  & 1234292944, 1236791704  & February-May  & 28.8  & 1.65, 1.33, 145.3 & 8.45 \\
		  Im. 8      & 1121507392, 1121593824  & 2015 &                1242049824, 1242221248  & 2019 &   & & \\
		       & 1121420968, 1121680256  &  &                1241705168, 1242479744  &  &   & & \\
        \hline
		Final image &  &  &  & & 3.8 hrs  & 1.5, 1.2, 155 & 4.0\\		  
		\hline
	\end{tabular}
\end{table*} 

\textbf{Data availability}\\

The raw visibility data from the MWA that support the findings of this study (as detailed in Table~\ref{table:1}) are publicly available from the MWA All-Sky Virtual Observatory; \url{https://asvo.mwatelescope.org/}. The MWA radio image data (as displayed in Figs.~\ref{fig:1} and \ref{fig:2} and Extended Data Figure~\ref{fig:3a}) can be accessed via the Aladin Sky Atlas; \url{https://aladin.u-strasbg.fr/}. Optical image data will be made available upon reasonable request to the corresponding author.\\

\textbf{Code availability}\\

No custom code or algorithm was developed as part of this work, apart from simple scripting routines written in the \sc python \rm language and used to run standard astronomy software tools, as described in the text. \\


\textbf{Corresponding author}\\

Correspondence and requests for materials should be addressed to B.M. \\ \\

\textbf{Acknowledgements}\\

This scientific work makes use of the Murchison Radio-astronomy Observatory, operated by CSIRO. We acknowledge the Wajarri Yamatji people as the traditional owners of the Observatory site. Support for the operation of the MWA is provided by the Australian Government (NCRIS), under a contract to Curtin University administered by Astronomy Australia Limited. We acknowledge the Pawsey Supercomputing Centre which is supported by the Western Australian and Australian Governments. MG acknowledges partial support by NASA Chandra GO8-19104X/GO9-20114X and HST GO-15890.020/023-A, and the \textit{BlackHoleWeather} program. MC and MM would like to thank St\'{e}phane Courteau and Doug Simons for CFHT Director's discretionary time via the Mauna Kea Graduate School program. Research by DC is supported by NSF grant AST-1814208. BM was supported by an ARC Future Fellowship awarded to Cathryn Trott under grant FT180100321. The authors would like to thank Annalisa Pillepich, Raffaella Morganti, Rolf Wahl-Olsen and Mike Sidonio for useful discussions and guidance during preparation of the manuscript. \\

\textbf{Author contributions}\\

Processing of the radio data was performed by B.M. A.R.O. contributed software development crucial to the success of the radio data processing. Processing of the lower resolution H$\alpha$ data in Fig.~\ref{fig:2} and visualisation of the radio data in Extended Data Figure~\ref{fig:3a} was performed by C.M. Processing of the CFHT data was performed by M.S.C., M.M. and S.D.J.G. R.P.K. provided the X-ray data and input regarding its role in the analysis. D.C. provided the RGB star map and input regarding the merger history of Centaurus~A. M.G., S.S.S., S.V., S.T. and J.B-H. provided input on modelling, theoretical aspects and consistency with observations. B.M.G. and M.J-H. played critical roles in the upgrade of the MWA that enabled the radio observations. All authors contributed to the interpretation of the data and the discussion presented, and all were involved in the writing of the paper.\\

\textbf{Competing interests}\\

The authors declare no competing interests.

\bsp	
\label{lastpage}
\end{document}